\documentclass[12pt]{article}
\usepackage{amsmath,amssymb,booktabs,hyperref,siunitx,array,geometry}
\usepackage[utf8]{inputenc}
\usepackage[T1]{fontenc}
\usepackage{graphicx}
\usepackage[square,numbers]{natbib}
\title{MAD-Accretion Scenario in High-Mass X-Ray Binaries with Neutron Stars}
\author{
    N. R. Ikhsanov\textsuperscript{1,2},
    M. Yu. Piotrovich\textsuperscript{1},
    N. G. Beskrovnaya\textsuperscript{1,3,*} \\
    \scriptsize\textsuperscript{1}Central (Pulkovo) Astronomical Observatory, Russian Academy of Sciences, St. Petersburg, 196140 Russia \\
    \scriptsize\textsuperscript{2}Institute of Applied Astronomy, Russian Academy of Sciences, St. Petersburg, 191187 Russia \\
    \scriptsize\textsuperscript{3}Special Astrophysical Observatory, Russian Academy of Sciences, Nizhnii Arkhyz, 369167 Russia\\
    \scriptsize\textsuperscript{*}E-mail: beskrovnaya@yahoo.com\\
    {\scriptsize \it Scheduled for publication in Astrophysical Bulletin, 2025, Vol. 80, No. 1, pp. 122–131.}
}

\begin{document}

\maketitle

\begin{abstract}
The wind-fed accretion of gas with an intrinsic magnetic field onto neutron stars in high-mass X-ray binaries (HMXBs) is discussed. The criteria for the implementation of quasi-spherical accretion in the system, the formation of a Keplerian accretion disk and a non-Keplerian magnetically arrested disk (MAD-accretion scenario) are formulated. It is shown that the period of the equilibrium rotation of a neutron star in the MAD-accretion scenario, in addition to the parameters of the binary system, is also determined by the velocity, temperature, and degree of magnetization of the stellar wind of its massive companion. Within the hypothesis of the equilibrium rotation, we have estimated the parameters of the stellar wind for the most fully-studied quasi-equilibrium accreting pulsars in the HMXBs corresponding to the criteria for the MAD-accretion scenario implementation. We show that the period of the equilibrium rotation of a neutron star in the system, all other things being equal, increases with the growth of the magnetization degree of the stellar wind component of its massive companion flowing out in the plane of the system's orbit. Therefore, the periods of accreting pulsars in systems with similar parameters can differ from each other by several orders of magnitude and, under favorable conditions, reach values of several thousand and even tens of thousands of seconds.

\end{abstract}

\section{Introduction}
Accreting pulsars in HMXBs are neutron stars that form close pairs with massive early-type stars. The X-ray emission from these objects is predominantly thermal and is caused by gas accreting onto the surface of a neutron star. The donor of the accretion flow is the massive component of the system, which in most cases does not fill its Roche lobe but intensively loses the matter in the form of the stellar wind. The gas captured by the neutron star from the wind forms an accretion flow which, moving towards the star, eventually reaches its surface in the region of the magnetic poles. Due to the axial rotation of the star, whose magnetic axis is inclined to its rotation axis, the intensity of X-ray emission detected by a remote observer experiences periodic variations creating the effect of an X-ray pulsar.

Modeling the process of the mass exchange between the components of the systems described above is usually carried out on the basis of the energy and time characteristics of their X-ray emission. The latter include, first of all, the pulsar period and the rate of its changing. Analysis of these characteristics allows us to estimate the moment of force applied to the neutron star by the accretion flow at the current stage of the HMXB evolution and to limit the number of possible scenarios for the matter flow in these systems.

From a theoretical point of view, the rotation period of an accreting neutron star tends over time to its equilibrium value, at which the total resulting moment of force applied to the star during its interaction with the accretion flow becomes zero. The value of this period within the framework of traditional accretion scenarios is estimated by the expressions (see, for example, \citet{ikhsanov2007} and the literature cited there):

\begin{equation}
P_{\text{eq}}^{(d)} \simeq 7\,\text{s}\,\mu_{30}^{6/7}\dot{M}_{16}^{-3/7}m^{-5/7},
\label{eq:keplerian}
\end{equation}

\noindent when accreting from the Keplerian disk, and

\begin{equation}
P_{\text{eq}}^{(s)} \simeq 60\,\text{s}\,\xi_{0.2}^{-1/2}\mu_{30}\dot{M}_{16}^{-1/2}m^{-3/2} \left(\frac{v_{\text{rel}}}{\SI{400}{km\,s^{-1}}}\right)^2 \left(\frac{P_{\text{orb}}}{\SI{10}{days}}\right)^{1/2},
\label{eq:quasispherical}
\end{equation}

\noindent when implementing a quasi-spherical accretion scenario in the system, respectively. Here, $\mu_{30}$ is a dipole magnetic moment ($\mu = (1/2)B_{\text{ns}}R_{\text{ns}}^{3}$) of a neutron star in units $\SI{e30}{G\,cm^3}$ and $m$ is its mass, $M_{\text{ns}}$, in units $1.4\, M_{\odot}$. This normalization assumes that the magnetic field strength on the surface of a neutron star, $B_{\text{ns}}$, with its radius $R_{\text{ns}} \sim \SI{10}{km}$ is $\SI{2e12}{B_{12}\,G}$ (here, $B_{12} = B_{\text{ns}}/\SI{e12}{G}$). The parameter $\dot{M}_{16}$ is the accretion rate in units $\SI{e16}{g\,s^{-1}}$, at which the luminosity of the accretion source, $L_x = \dot{M}GM_{\text{ns}}/R_{\text{ns}}$, reaches the value $L_x = \SI{e36}{L_{36}\,erg\,s^{-1}}$ (where $L_{36} = L_x/\SI{e36}{erg\,s^{-1}}$); $v_{\text{rel}}$ is the velocity of a neutron star relative to the wind of its massive companion, the adopted normalization of which we discuss in Section 2.2; $P_{\text{orb}}$ is the orbital period of the binary system and $\xi_{0.2} = \xi/0.2$ is the dimensionless parameter taking into account the dissipation of the angular momentum in a quasi-spherical accretion flow due to density and velocity gradients, normalized in accordance with the results of numerical modeling \citep{ruifert1999}. The normalizations of the remaining parameters adopted in these expressions correspond to the values of the HMXB parameters presented in Table \ref{tab:hmxb_params}.

\begin{table}[ht!]
\small
\centering
\caption{Parameters of HMXBs (see \citet{cox2005, finger2010, falanga2015, hambaryan2022, kim2023} and references therein). The average luminosity in the X-ray range $L_{36} = L_{\rm x}/(10^{36}\,\text{erg/s})$, the magnetic field of a neutron star $B_{12} = B_{\rm ns}/(10^{12}\text{G})$. For 4U~1907+09 and 4U~2206+54, the canonical mass value of a neutron star is given}
\label{tab:hmxb_params}
\begin{tabular}{@{}lccccccccc@{}}
\toprule
HMXB & $P_s$, s & $P_{\text{orb}}$, d & $M_{\text{ns}}$ ($M_\odot$) & $L_{36}$ & $B_{12}$ & $d$, kpc & Sp. class & $M_{\rm opt},\ M_\odot$\\
\midrule
 OAO 1657-415 &   37 &  10.5 & 1.74 &  3.0 & 3.2 & 6.4 & Ofpe/WN9 & 17.5 \\
Vela      X-1 &  283 &   9.0 & 2.12 &  4.0 & 2.6 & 2.0 & B0.5  Ib & 26.0 \\
  4U  1907+09 &  441 &   8.0 & 1.40 &  2.0 & 2.1 & 4.0 & O8-9  Ia & 27.0 \\
  4U  1538-52 &  525 &   4.0 & 1.02 &  2.0 & 2.3 & 4.5 &   B0 Iab & 16.0 \\
  4U  2206+54 & 5554 &  19.2 & 1.40 &  0.3 & 4.0 & 2.6 & O9.5 V & 23.5 \\
\bottomrule
\end{tabular}
\end{table}

The spin evolution of pulsars, monitored for several decades, supports the hypothesis of the equilibrium rotation of neutron stars in these systems. However, the periods of most of the currently known accreting pulsars, distributed over a wide range of values from several seconds to tens of thousands of seconds, differ significantly from the above estimates of the equilibrium period and do not show a stable correlation with the parameters of binary systems (see, e.g., \citet{kim2023} and references therein). Attempts to explain this discrepancy by inaccuracies in the estimates of the magnetic field of neutron stars (see, e.g., \citet{shakura1975,klus2014}) in the case of the pulsars considered in this paper are untenable, since the values of the magnetic field on the surface of neutron stars in these systems were estimated from observations of cyclotron lines in their X-ray spectrum (see Table \ref{tab:hmxb_params} and the literature cited there). The most striking example of the above contradiction is the HMXB 4U 2206+54, the rotation period of the neutron star in which exceeds \SI{5500}{s} and at the same time demonstrates a tendency to a stable increase. Within the framework of traditional scenarios, such behavior of the neutron star could only be characteristic of a magnetar. However, the magnetic field of the neutron star, measured from observations of the cyclotron line in the X-ray spectrum of this system, does not exceed $\SI{4e12}{G}$ (for a more detailed discussion, see, e.g., \citet{ikhsanov2013} and references therein).

Another possible reason for the above discrepancy may be the oversimplification of traditional accretion scenarios leading to inaccuracies in estimating the equilibrium period of accreting pulsars. This possibility was analyzed, in particular, in the paper by \citet{ikhsanov2015}, where the process of mass exchange between the components of an HMXB was considered within the framework of the MAD-accretion scenario\footnote{The authors of this paper used the term "magnetically levitated accretion", although, the scenario they considered is identical to MAD accretion.}. This scenario, first proposed in the papers\footnote{The scenario they proposed received its current name "Magnetically Arrested Disk" (MAD) only 29 years later in the paper by \citet{narayan2003}.} by \citet{bisnovatyi1974,bisnovatyi1976}, includes an additional factor, which is the intrinsic magnetic field of the accretion flow. Under the influence of this factor, under certain conditions, a magnetic non-Keplerian disk can form in the system, the movement of gas in which is completely controlled by its own magnetic field \citep{bisnovatyi2019}. The value of the equilibrium period of the neutron star in this case differs significantly from the values expected in traditional accretion scenarios, and, in addition to the parameters of the binary system, is also determined by the physical conditions in the stellar wind of its massive companion. We briefly describe the general picture of the accretion realized in this scenario in Section 2. Following the basic principles of this model, in Section 3 we present estimates of the stellar wind parameters of the five most thoroughly studied HMXBs with similar main parameters. We briefly discuss the key features of the scenario we present in Section 4, and draw the main conclusions of our study in Section 5.

\section{MAD-Accretion Scenario in HMXBs: the Overall Picture}

We consider HMXBs with the wind-fed accretion onto a neutron star with a strong magnetic field. This implies that the massive component of the system does not fill its Roche lobe and loses matter in the form of a stellar wind. A neutron star with the mass $M_{\text{ns}}$, moving through the wind of its companion, captures gas located at the distance $r \leq r_\text{G}$ from its center and accretes it onto its surface. Here,

\begin{equation}
r_\text{G} = \frac{2GM_{\text{ns}}}{v_{\text{rel}}^2}
\label{eq:bondi}
\end{equation}

determines the radius of the gravitational capture of a neutron star (the Bondi radius).

The captured gas has its own magnetic field, the initial source of which is the magnetic field of the stellar wind. The characteristic scale of the magnetic field in the region of gas capture by a neutron star (at the Bondi radius) is comparable in order of magnitude to the size of the binary system and for the parameters of interest to us significantly exceeds the Bondi radius. This allows us to consider the accretion process in the approximation of a large-scale uniform magnetic field and to normalize the initial magnetic field strength in the accretion flow, $B_{f0} = B_f(r_\text{G})$, to the magnetic field strength in the stellar wind of a massive star at a distance of about the size of a binary system.

A distinctive feature of the implementation of the MAD-accretion scenario in HMXBs is the presence of initial angular momentum in the gas that the neutron star captures from the stellar wind of its companion at the Bondi radius. The source of the angular momentum of the accretion flow is the orbital rotation of the binary system. The initial angular velocity of the gas captured at the Bondi radius, $\Omega_{f0} = \Omega_f(r_\text{G})$, as a result of this, turns out to be of the order of the orbital angular velocity of the system, $\Omega_{\text{orb}} = 2\pi/P_{\text{orb}}$.

The MAD-accretion scenario in HMXBs can be roughly represented as a sequence of three main phases:
\begin{enumerate}
\item the quasi-spherical gas accretion with the angular momentum and intrinsic magnetic field;
\item a slowly rotating (non-Keplerian) magnetically arrested disk interacting with the magnetic field of a neutron star;
\item the accretion inside the magnetosphere of a neutron star, during which the gas moving from the boundary of the magnetosphere along the magnetic field lines of the neutron star, reaches its surface in the region of the magnetic poles.
\end{enumerate}

\subsection{Quasi-Spherical Accretion Phase}
The gas captured by a neutron star at the Bondi radius, under the action of its gravitational field, forms an accretion flow. Provided that $v_{\text{rel}} > v_{\text{orb}}$, where

\begin{equation}
v_{\text{orb}} = \left(\frac{2\pi GM_{\text{opt}}}{P_{\text{orb}}}\right)^{1/3}
\label{eq:vorb}
\end{equation}

determines the velocity of the neutron star in a circular orbit around its massive companion, the captured gas moves towards the neutron star initially in free fall in the form of a quasi-spherical flow. As it approaches the neutron star, the dynamic pressure of the flow in the radial,

\begin{equation}
p_{\text{ram}}(r) = \rho(r)v_{\text{ff}}^{2}(r) \propto r^{-5/2},
\label{eq:pram}
\end{equation}

and azimuthal,

\begin{equation}
p_\phi = \rho(r)v_\phi^{2}(r) \propto r^{-7/2},
\label{eq:pphi}
\end{equation}

directions increases and its own magnetic field grows leading to an increase of the magnetic pressure in the accretion flow,

\begin{equation}
p_m(r) = \frac{B_f^2(r)}{4\pi} \propto r^{-4}.
\label{eq:pm}
\end{equation}

Here, $\rho(r) = \frac{\dot{M}}{4\pi r^2 v_{\text{ff}}(r)}$ is the gas density in a spherical accretion flow approaching a neutron star at free fall velocity $v_{\text{ff}}(r) = (2GM_{\text{ns}}/r)^{1/2}$, and $\dot{M}$ is the accretion rate, the value of which in the stationary picture of accretion from the Bondi radius to the surface of a neutron star is assumed to be constant. The parameter $v_\phi(r) = r \Omega_f(r)$ is the linear velocity of the gas in the azimuthal direction and

\begin{equation}
\Omega_f(r) = \xi \Omega_{f0} \left(\frac{r_\text{G}}{r}\right)^2
\label{eq:omegaf}
\end{equation}

is its angular velocity, the change in which, as the accretion flow approaches a neutron star, occurs due to the angular momentum conservation. Finally,

\begin{equation}
B_f(r) = B_{f0} \left(\frac{r_\text{G}}{r}\right)^2
\label{eq:bf}
\end{equation}

determines the intrinsic magnetic field in the gas at the spherical accretion stage, the change of which, as the gas approaches the star, occurs due to the conservation of the magnetic flow.

The above factors, along with the neutron star parameters, determine the accretion scenario realized in HMXBs. In particular, the shortest possible distance, to which the infalling gas is able to approach the neutron star in the form of a quasi-spherical flow, is determined by the Alfven radius\footnote{It should be noted that the radius of the neutron star magnetosphere, in general, differs from the Alfven radius, which has the physical meaning of the equilibrium radius, at which a spherical accretion flow is stopped by the dipole magnetic field of a neutron star \citep{beskrovnaya2024}.},

\begin{equation}
r_\text{A} = \left(\frac{\mu^2}{\dot{M}(2GM_{\text{ns}})^{1/2}}\right)^{2/7},
\label{eq:ralfven}
\end{equation}

which is the solution of the balance equation between the dynamic pressure of the spherical flow and the pressure of the dipole magnetic field of a neutron star. Here, $\mu = (1/2)B_{\text{ns}}R_{\text{ns}}^{3}$ is the dipole magnetic moment of a neutron star with the radius $R_{\text{ns}}$, the magnetic field strength on the surface of which is $B_{\text{ns}}$.

There are, however, at least two additional factors that contribute to the transformation of a quasi-spherical flow into a disk at distances exceeding the Alfven radius. The first of these is the proper angular momentum of the captured gas, which contributes to the accretion-driven spin-up of the flow. As the angular velocity of the gas during accretion reaches the Keplerian velocity, the centrifugal force affecting the infalling matter becomes equal to the gravitational force of the star. The centrifugal barrier that arises in this case prevents further radial motion of the gas toward the star and results in the transformation of the original quasi-spherical flow into a Keplerian disk, in which the radial motion of the gas is determined by the viscosity of the disk contributing to the removal of the excess angular momentum from its inner to outer radius. The distance, at which this situation can be realized in HMXBs,

\begin{equation}
r_{\text{circ}} = \frac{\xi^2 \Omega_{\text{orb}}^2 r_\text{G}^4}{GM_{\text{ns}}},
\label{eq:rcirc}
\end{equation}

is called the circularization radius. Substituting expression (\ref{eq:bondi}) into (\ref{eq:rcirc}) and solving the inequality $r_\text{A} < r_{\text{circ}}$ relative to $v_{\text{rel}}$, we arrive at a necessary (but insufficient) condition for the formation of a Keplerian viscous disk in HMXBs, $v_{\text{rel}} < v_0$, where

\begin{equation}
v_0 \simeq \SI{400}{km\,s^{-1}} \xi_{0.2}^{1/4} m^{11/28} \mu_{30}^{-1/14} \dot{M}_{16}^{1/28} \left(\frac{P_{\text{orb}}}{\SI{10}{days}}\right)^{-1/4}.
\label{eq:v0}
\end{equation}

The second factor is the magnetic field of the flow itself which grows as the gas falls onto a neutron star. This leads to an increase of the magnetic pressure in the quasi-spherical flow, which will reach the value of its dynamic pressure at the radius

\begin{equation}
r_{\text{ml}} = \beta_0^{-2/3} r_\text{G} \left(\frac{c_{\text{so}}}{v_{\text{rel}}}\right)^{4/3} \simeq r_\text{G} \left(\frac{v_{\text{A0}}}{v_{\text{rel}}}\right)^{4/3}.
\label{eq:rml}
\end{equation}

Here, $\beta_0 = \beta(r_\text{G}) = 8\pi \rho_0 c_{\text{so}}^2 / B_{f0}^2$ is the ratio of the thermal and magnetic pressure, $c_{\text{so}} = c_s(r_\text{G})$ is the sound speed, and $v_{\text{A0}} = B_{f0} / \sqrt{4\pi \rho(r_\text{G})}$ is the Alfven velocity in the accretion flow at the Bondi radius.

Under the influence of this factor, the movement of gas in the space region $r \leq r_{\text{ml}}$ switches from the free-fall state to the state of diffusion, whose parameters are determined by the efficiency of dissipation of the magnetic field in the accretion flow. This conclusion, noted even at the early stages of the accretion theory development \citep{shvartsman1971}, turns out to be quite natural from a physical point of view, if we take into account that the volume occupied by gas in a spherical accretion flow, as it approaches a neutron star, rapidly decreases. As a result, the flow of matter in the free-fall state is possible only in the absence of factors capable of preventing a decrease in the volume of the flow. The presence of a magnetic field in the flow violates this condition creating an excess of the magnetic pressure\footnote{It should be noted that the magnetic field pressure is not isotropic and does not affect significantly the gas motion along its field lines.} in the region $r \leq r_{\text{ml}}$, which prevents further decrease of the spherical flow volume. Under this condition, the gas motion proceeds in the diffusion regime, the parameters of which are determined by the dissipation rate of the magnetic field in the accretion flow \citep{ikhsanov2012}.

Analyzing the structure of the matter flow in the space region $r \leq r_{\text{ml}}$, first of all, it should be noted that the initial magnetic field of the flow does not have spherical symmetry and its configuration also undergoes significant changes during the accretion process. While the perpendicular scale in a spherical accretion flow decreases, $l_\perp \propto r^{-2}$, its longitudinal scale increases, $l_\parallel \propto r^{1/2}$ \citep{bisnovatyi1970}. This results in the fact that the intrinsic magnetic field of the flow during the accretion process acquires the configuration of the disk, the plane of which is located perpendicular to the initial direction of the magnetic field in the gas at the distance $r \geq r_\text{G}$ \citep{bisnovatyi1974}. It should also be noted that the magnetic field pressure is not isotropic and does not affect significantly the gas motion along its field lines. As a result, the gas flow in the space region $r \leq r_{\text{ml}}$ will occur predominantly along the field lines of the flow's own magnetic field repeating its configuration and forming a magnetized disk.

The results of numerical modeling of the spherical gas accretion with account for the magnetic field of the flow itself, first presented in the papers by \citet{igumenshchev2003} and \citet{narayan2003}, generally confirmed the conclusions of the above-described scenario with the formation of a magnetic disk configuration called magnetically arrested disk (see \citet{bisnovatyi2019} and references therein).

\subsection{Conditions for the Implementation of the MAD-Accretion Scenario in HMXBs}
The matter movement towards a neutron star in HMXBs with the wind-fed accretion can occur according to the following main scenarios:
\begin{itemize}
\item the quasi-spherical accretion;
\item the accretion from a viscous Keplerian disk;
\item the MAD-accretion.
\end{itemize}

The quasi-spherical accretion scenario is realized in HMXBs under the condition
\begin{equation}
r_\text{A} > \max\{r_{\text{circ}}, r_{\text{ml}}\},
\label{eq:quasi_condition}
\end{equation}
implying that the infall of gas in the quasi-spherical flow will be stopped by the magnetic field of a neutron star before its spin-up and magnetization reach values sufficient to transform the flow into a viscous or magnetized disk, respectively. Solving this inequality with respect to $v_{\text{rel}}$, we arrive at the condition $v_{\text{rel}} > \max\{v_{\text{ma}}, v_0\}$, where
\begin{equation}
v_{\text{ma}} \approx \SI{2380}{km\,s^{-1}} \beta_0^{-1/5} \mu_{30}^{6/35} m^{12/35} \dot{M}_{16}^{3/35} T_7^{1/5}
\label{eq:vma}
\end{equation}
is a solution to the equation $r_\text{A} = r_{\text{ml}}$, and $v_0$ is a solution to the equation $r_\text{A} = r_{\text{circ}}$, the value of which is represented by expression (\ref{eq:v0}). The temperature of the stellar wind in the region of the neutron star's orbit\footnote{Let us note that the systems we are considering are close, and the radius of the neutron star's orbit in most of them does not exceed two radii of its massive companion.}, $T_w$, in this expression is normalized to the average value of the corona temperature of massive stars of early spectral types, $T_7 = T_w / \SI{e7}{K}$, obtained from observations of these objects in the X-ray range \citep{schulz2003}.

The formation of a Keplerian disk in HMXBs is possible only under the condition
\begin{equation}
r_{\text{circ}} > \max\{r_\text{A}, r_{\text{ml}}\},
\label{eq:kepler_condition}
\end{equation}
which corresponds to the fulfillment of the inequality $v_{\text{rel}} < \min\{v_{\text{ca}}, v_0\}$, where
\begin{equation}
v_{\text{ca}} \approx \SI{57}{km\,s^{-1}} \xi_{0.2}^{3/7} \beta_0^{1/7} m^{3/7} P_{10}^{-3/7} T_7^{-1/7}
\label{eq:vca}
\end{equation}
is the solution of the equation $r_{\text{circ}} = r_{\text{ml}}$.

And finally, the MAD-accretion scenario is implemented in the system under the condition
\begin{equation}
r_{\text{ml}} > \max\{r_\text{A}, r_{\text{circ}}\},
\label{eq:mad_condition}
\end{equation}
which is fulfilled if the velocity of a neutron star relative to the stellar wind of its massive companion is in the range of
\begin{equation}
v_{\text{ca}} < v_{\text{rel}} < v_{\text{ma}}.
\label{eq:velocity_range}
\end{equation}

The degree of magnetization of the stellar wind required to implement the MAD-accretion scenario can be found by solving the inequality $v_{\text{ca}} < v_{\text{ma}}$ for the parameter $\beta_0$, which brings us to the condition $\beta_0 < \beta_{\text{max}}$, where
\begin{equation}
\beta_{\text{max}} \approx 10^4 \xi_{0.2}^{-5/4} \mu_{30}^{-1/2} \dot{M}_{16}^{1/4} m^{-1/4} P_{10}^{5/4} T_7
\label{eq:betamax}
\end{equation}
is the solution of the equation $v_{\text{ca}} = v_{\text{ma}}$.

This condition corresponds to the magnetic field strength in the stellar wind of a massive star at a distance of the order of the system size $B_{f0} > B_{\text{min}}$, where
\begin{equation}
B_{\text{min}} \approx \SI{50}{mG} \xi_{0.2}^{5/8} \mu_{30}^{1/4} \dot{M}_{16}^{3/8} m^{-7/8} \left(\frac{P_{\text{orb}}}{\SI{10}{days}}\right)^{-5/8} \left(\frac{v_{\text{rel}}}{\SI{400}{km\,s^{-1}}}\right)^{3/2}.
\label{eq:bmin}
\end{equation}

The velocity of a neutron star relative to the stellar wind of its massive companion, $v_{\text{rel}}$, is here and below normalized to the value $\SI{400}{km\,s^{-1}}$, corresponding to the sound speed in the gas heated to a temperature of $\SI{e7}{K}$, which, in turn, corresponds to the normalization of the stellar wind temperature adopted by us on the basis of X-ray observations of the coronae of early-type massive stars (see above).

\subsection{Equilibrium Spin Period of a Neutron Star}
One of the features of the accretion process in HMXBs is the presence of the initial angular momentum of the accretion flow. As a result, the magnetized disk formed in the system within the MAD-accretion scenario must have axial rotation. The angular velocity of the gas in the disk formation region determined by the expression
\begin{equation}
\Omega_f(r_\text{ml}) = \xi \Omega_\text{orb} \left(\frac{r_\text{G}}{r_\text{ml}}\right)^2,
\label{eq:omegaf_rml}
\end{equation}
corresponds to the angular velocity of the gas, to which it was spun-up in a quasi-spherical flow before its transformation into a magnetically arrested disk (MAD) at the radius $r_{\text{ml}}$.

The issue of the radial distribution of the angular velocity of matter in a magnetized disk is complex and remains poorly studied to date. It cannot be ruled out that the rotation of the disk, in which the motion of matter is completely controlled by its own magnetic field, remains close to that of a solid body. Otherwise, differential rotation would lead to the generation of a toroidal component of the magnetic field and, thus, to an increase in the magnetic energy, which, in the situation under consideration, initially dominates the energy balance of the disk. Taking this argument into account, in this paper we will limit ourselves to discussing the case of the solid-body rotation of the disk with the angular velocity $\Omega_f(r_{\text{ml}})$ represented by expression (\ref{eq:omegaf_rml}).

The interaction of the disk with the magnetic field of a neutron star leads to the formation of its magnetosphere. At the boundary of the magnetosphere, corresponding to the inner radius of the disk, the accretion flow diffuses into a magnetic field of the neutron star and, moving along its field lines, reaches the surface of the star in the region of the magnetic poles. The radius of the magnetosphere $r_m$ and, accordingly, the inner radius of the disk are estimated from the continuity equation, in which the accretion rate is assumed to be equal to the rate of gas diffusion from the inner radius of the disk into the magnetic field of a neutron star. An auxiliary condition that allows us to determine the density of the accretion flow at the magnetosphere boundary implements the balance between the pressure of the dipole magnetic field of a neutron star and the pressure exerted by the accretion disk\footnote{It should be noted that the radius of the neutron star magnetosphere, in general, differs from the Alfven radius, which has the physical meaning of the equilibrium radius, at which a spherical accretion flow is stopped by the dipole magnetic field of a neutron star \citep{beskrovnaya2024}.}. Finally, normalizing the diffusion coefficient of the accretion flow into the magnetic field of the neutron star by the Bohm diffusion coefficient (determined by the processes of reconnection of the field lines of the magnetosphere and disk field, as well as by the instabilities of the drift-dissipative type at the magnetosphere boundary), we arrive at the following estimate of the magnetosphere radius: $r_m \geq r_N$, where \citep{beskrovnaya2024}
\begin{equation}
r_N \approx \SI{2.3e7}{cm} \lambda_0 \mu_{30}^{6/11} \dot{M}_{16}^{-4/11} m^{-1/11}
\label{eq:rn}
\end{equation}
and $\lambda_0$ is the dimensionless parameter, whose value within the framework of our assumptions is of the order of unity.

The equation of the equilibrium rotation of a neutron star, the rotation axis of which coincides with the rotation axis of the disk, in the situation we are considering, can be written as
\begin{equation}
\dot{M} \left( r_m^2 \Omega_f - R_{\text{ns}}^2 \omega_s \sin^2 \theta \right) - \Delta_m = 0,
\label{eq:equilibrium}
\end{equation}
where $\omega_s = 2\pi / P_s$ is the angular velocity of a neutron star, $\theta$ is the angle between the rotation axis of the neutron star and the axis of its magnetic field, and $\Delta_m$ is the absolute value of the moment of force arising from the interaction of the accretion flow with the magnetic field of the star and leading to partial dissipation of the angular momentum of the gas accreted from the inner radius of the disk onto its surface.

The first term in brackets on the left side of this equation, $J(f) = r_m^2 \Omega_f$, determines the specific angular momentum of the gas located at the inner radius of the disk, $r_m$, and rotating with the angular velocity $\Omega_f$. The second term in brackets, $J(\text{ns}) = R_{\text{ns}}^2 \omega_s \sin^2 \theta$, determines the specific angular momentum of the gas that has already reached the surface of a neutron star in the region of its magnetic poles and rotates with it at the angular velocity $\omega_s$. The difference between these values, multiplied by the mass of gas accreted by the star per unit time onto its surface from the inner radius of the disk, corresponds to the moment of force applied to the neutron star by the accretion flow with an accuracy of up to the dissipative term $\Delta_m$.

As an illustration, we can first consider the simplest case, in which we assume that the rotation axis and the magnetic axis of a neutron star are perpendicular (the case of an orthogonal rotator, $\sin^2 \theta = 1$), and neglect the effects of dissipation ($\Delta_m = 0$). Solving equation (\ref{eq:equilibrium}) under these conditions, we arrive at the value of the angular velocity of a neutron star
\begin{equation}
\omega_s^{(eq)} \approx \Omega_f (r_{\text{ml}}) \left( \frac{r_m}{R_{\text{ns}}} \right)^2,
\label{eq:omega_eq}
\end{equation}
at which its rotation is in equilibrium. In our simplified example, this state corresponds to the equality $J(\text{ns}) = J(f)$, with which the accretion of gas from the inner radius of the disk onto the stellar surface does not lead to a change in the angular momentum of the latter. In Section 3, we estimate the parameters of the stellar wind of the massive component of an HMXB, at which the implementation of such a condition is possible using a sample of the most fully-studied quasi-equilibrium accreting pulsars in galactic HMXBs with similar orbital periods.

\section{Stellar Wind Parameters}
Within the accretion picture described in Section 2, we consider an HMXB in which the scenario of MAD-accretion onto a rotating magnetized neutron star is realized. The mass exchange between the components of the system leads to the formation of a solidly rotating MAD, whose angular velocity, estimated with expression (\ref{eq:omegaf_rml}), is significantly smaller than the Keplerian velocity. The outer radius of the disk, $r_{\text{ml}}$, represented by expression (\ref{eq:rml}), does not exceed the radius of gravitational capture of a neutron star, $r_\text{G}$, determined by expression (\ref{eq:bondi}). The minimum possible value of the inner radius of the disk, $r_N$, is estimated by expression (\ref{eq:rn}) and for the parameters of the pulsars under study, it significantly exceeds the radius of the neutron star. In the process of accretion of matter from the disk onto the surface of a neutron star, the angular velocity of its axial rotation evolves toward a value estimated by expression (\ref{eq:omega_eq}), at which the moment of force applied to a neutron star by the accretion flow turns to zero. The degree of magnetization of the stellar wind, at which this condition is achieved can be estimated by combining expressions (\ref{eq:bondi}), (\ref{eq:rml}), (\ref{eq:rn}), (\ref{eq:omegaf_rml}), (\ref{eq:omega_eq}) and solving the resulting equation with respect to $\beta_0$. Using the normalization of parameters we have adopted, we find
\begin{equation}
\beta_0 \simeq 50 \xi_{0.2}^{-3/4} \mu_{30}^{-9/11} \dot{M}_{16}^{6/11} R_6^{3/2} m^{3/22} T_7 \left( \frac{P_{\text{orb}}}{\SI{10}{days}} \right)^{3/4} \left( \frac{P_s}{\SI{40}{s}} \right)^{-3/4} \left( \frac{v_{\text{rel}}}{\SI{400}{km\,s^{-1}}} \right)^{-2},
\label{eq:beta0}
\end{equation}
where $R_6 = R_{\text{ns}} / \SI{e6}{cm}$.

The values of $\beta_0$ expected in the case of the MAD-accretion scenario in the HMXBs we studied are given in Table \ref{tab:results} together with estimates of the minimum possible radius of the neutron star magnetosphere, $r_N$, and the Alfven radius, $r_\text{A}$. The same table shows the minimum, $v_{\text{ca}}$, and maximum, $v_{\text{ma}}$, values of the relative velocity, at which the MAD-accretion scenario is realized in the system, together with the average estimate of the orbital velocity of a neutron star, $v_{\text{orb}}$ and the lower limit of the magnetic field in the stellar wind at the radius of the neutron star's orbit, $B_{\text{min}}$, which determines the threshold of applicability of the accretion scenario under consideration.

\begin{table}[ht!]
\footnotesize
\centering
\caption{Results of model calculations for $v_{\text{rel}} = \SI{400}{km\,s^{-1}}$ and $T_w = \SI{e7}{K}$}
\label{tab:results}
\begingroup
\setlength{\tabcolsep}{3pt}
\begin{tabular}{@{}lcccccccccc@{}}
\toprule
HMXBs & $P_{\text{orb}}$, d & $P_s$, s & $\beta_0$ & $r_N$, $\SI{e8}{cm}$ & $r_\text{A}$, $\SI{e8}{cm}$ & $v_{\text{ca}}$, $\SI{}{km/s}$ & $v_{\text{orb}}$, $\SI{}{km/s}$ & $v_{\text{ma}}$, $\SI{}{km/s}$ & $B_{\text{min}}$, mG \\
\midrule
OAO 1657-415 & 10.5 & 37 & 44.0 & 0.3 & 7.0 & 105 & 253 & 1134 & 31 \\
Vela X-1 & 9.0 & 283 & 10.9 & 0.2 & 5.9 & 100 & 304 & 1675 & 28 \\
4U 1907+09 & 8.0 & 441 & 6.9 & 0.2 & 6.0 & 82 & 320 & 1612 & 37 \\
4U 1538-52 & 4.0 & 525 & 3.8 & 0.2 & 6.1 & 89 & 338 & 1648 & 87 \\
4U 2206+54 & 19.2 & 5555 & 0.4 & 0.7 & 14.9 & 38 & 228 & 2149 & 12 \\
\bottomrule
\end{tabular}
\endgroup
\end{table}

As can be seen from this table, the period of equilibrium rotation of a neutron star in an HMXB with the MAD-accretion, all other things being equal, increases with the degree of magnetization of the stellar wind component flowing out in the plane of the system's orbit. In the case of long-period pulsars, whose period reaches several thousand seconds, the upper limit on the value of the plasma parameter $\beta_0$, corresponding to the equilibrium rotation of a neutron star in the system, turns out to be of the order of unity, which corresponds to the balance between the magnetic and gas pressures in the stellar wind of a massive star.

Estimates of the lower, $v_{\text{ca}}$, and upper, $v_{\text{ma}}$, limits of the relative velocity $v_{\text{rel}}$, included in condition (\ref{eq:velocity_range}), indicate the fundamental possibility of implementing the MAD-accretion scenario in HMXBs presented in Table \ref{tab:results}. Indeed, the velocity $v_{\text{ca}}$ for all the systems we considered in a wide range of acceptable parameters does not exceed the value of the orbital velocity of the neutron star, which, in turn, determines the lower limit of the relative velocity of a neutron star in an HMXB.

On the other hand, there are good reasons to believe that the value of the relative velocity, $v_{\text{rel}}$, in the HMXBs we consider is significantly smaller than the values of $v_{\text{ma}}$ given in Table \ref{tab:results}. We come to this conclusion, in particular, by comparing the rate of gas capture by the neutron star from the wind of its massive companion,
\begin{equation}
\dot{M}_c = \pi r_\text{G}^2 \rho_0 v_{\text{rel}} = \frac{4\pi (GM_{\text{ns}})^2 \rho_0}{v_{\text{rel}}^3},
\label{eq:mdot_c}
\end{equation}
with the rate of the gas accretion onto the surface of a neutron star, $\dot{M} = L_x R_{\text{ns}} / GM_{\text{ns}}$, estimated from the X-ray luminosity of the source $L_x$. Here, $\rho_0 \simeq \kappa_w \dot{M}_{\text{out}} / 4\pi a^2 v_{\text{rel}}$ determines the density of the stellar wind at a distance of the order of the size of a binary system, $a$, in the region of its interaction with a neutron star, $\dot{M}_{\text{out}}$ is the mass-loss rate of the massive component in the form of stellar wind and $\kappa_w$ is the factor that takes into account the anisotropy of the stellar wind, the value of which in the case under consideration is of the order of unity. Assuming that the only source of the accretion flow in HMXBs is the stellar wind of the massive component, we obtain the condition $\dot{M} \leq \dot{M}_c$, solving which, relative to $v_{\text{rel}}$, we eventually arrive at the inequality $v_{\text{rel}} \leq v_{\text{cr}}$, where (see, e.g., \citet{delgado2001,ikhsanov2024})
\begin{equation}
v_{\text{cr}} \simeq \SI{440}{km\,s^{-1}} \kappa_w L_{36}^{-1/4} R_6^{-1/4} m^{3/4} \left( \frac{\dot{M}_{\text{out}}}{\SI{e-7}{M_\odot\,yr^{-1}}} \right)^{1/4} \left( \frac{a}{\SI{0.25}{au}} \right)^{-1/2}
\label{eq:vcr}
\end{equation}
is a solution to the equation $\dot{M}_c(v_{\text{cr}}) = \dot{M}$. The obtained estimate of the critical value, $v_{\text{cr}}$, corresponds to the normalization of relative velocity adopted in this paper, and is significantly smaller than the values $v_{\text{ma}}$, calculated for the HMXBs parameters presented in Tables \ref{tab:hmxb_params} and \ref{tab:results}. It should also be noted that the estimated velocity of the stellar wind flowing out in the plane of the system's orbit is significantly lower than the typical value of the terminal wind velocity of early spectral-type stars. This may indicate an anisotropic nature of the stellar wind of these objects, where, along with the fast polar component, there is a relatively slow wind component flowing out near the plane of the binary system.

\section{Discussion}
A distinctive feature of the MAD-accretion is the variable nature and wide range of possible values of the rotation angular velocity of the magnetized disk, the boundaries of which, in addition to the parameters of the HMXBs, are determined by a combination of the velocity, temperature, and magnetic field of the stellar wind of the massive component (see expressions (\ref{eq:rml}) and (\ref{eq:omegaf_rml})). The non-stationarity of the stellar wind, as a result, leads to variations in the angular velocity of the disk and, ultimately, to changes in the period of the X-ray pulsar. The diversity of possible manifestations of the above-mentioned non-stationarity is also noteworthy. In particular, changes in the pulsar period caused by variations in the stellar wind velocity will correlate with changes in its X-ray luminosity determined by the rate, at which the neutron star captures gas from the wind of its companion (see expression (\ref{eq:mdot_c})). Variations in the parameters of the magnetic field of the stellar wind (strength and/or configuration) also lead to changes in the angular velocity of the disk and, accordingly, the period of the pulsar, which in this case will, however, occur without changing the X-ray luminosity of the source.

Estimates of the angular velocity of the accretion flow expected for the HMXBs under consideration in different accretion scenarios are given in Table \ref{tab:angular}. The first column of this table presents the value of the rotation angular velocity of a neutron star, $\omega_s=2\pi/P_s$, calculated from the measured period of the X-ray pulsations of the HMXBs. The second column presents the value of the ratio of the Keplerian angular velocity at the Alfven radius, $\Omega_{\text{(k)}}(r_\text{A})=\sqrt{GM_{\text{ns}}/r_\text{A}^3}$, to the angular velocity of a neutron star. The third column presents the ratio of the angular velocity of the quasi-spherical flow at the Alfven radius corrected for the angular momentum dissipation factor, $\xi_{0.2}\Omega_{\text{(sp)}}(r_\text{A})=0.2\,\Omega_{\text{orb}}(r_\text{G}/r_\text{A})^2$, to the angular velocity of the neutron star. The last column of Table \ref{tab:angular} gives the ratio of the rotation angular velocity of the gas at the inner radius of the magnetized disk, $r_N$, to the angular velocity of the neutron star, where the rotation angular velocity of the magnetized disk,
\begin{equation}
\Omega_{\text{(MAD)}}(r_N)=\omega_s\left(\frac{R_{\text{ns}}}{r_N}\right)^2,
\label{eq:omega_mad}
\end{equation}
is evaluated based on condition (\ref{eq:omega_eq}).

\begin{table}[ht!]
\centering
\caption{Estimations of the ratio of the angular velocity of the accretion flow at the magnetosphere boundary to the angular velocity of a neutron star for the accretion from a Keplerian disk, the quasi-spherical accretion, and in the MAD-accretion scenario}
\label{tab:angular}
\begin{tabular}{@{}lcccc@{}}
\toprule
HMXBs & $\omega_s$, $\text{rad}\,\text{s}^{-1}$ & $[\Omega_{\text{(k)}}(r_\text{A})/\omega_s]$ & $[\xi_{0.2}\Omega_{\text{(sp)}}(r_\text{A})/\omega_s]$ & $[\Omega_{\text{(MAD)}}(r_N)/\omega_s]$ \\
\midrule
OAO 1657-415 & 0.170 & 4.5 & 1.1 & $1.1 \times 10^{-3}$ \\
Vela X-1 & 0.022 & 45 & 13 & $2.5 \times 10^{-2}$ \\
4U 1907+09 & 0.014 & 69 & 23 & $2.5 \times 10^{-2}$ \\
4U 1538-52 & 0.012 & 78 & 50 & $2.5 \times 10^{-2}$ \\
4U 2206+54 & 0.001 & 246 & 21 & $2 \times 10^{-4}$ \\
\bottomrule
\end{tabular}
\end{table}

As can be seen from this Table \ref{tab:angular}, the angular velocity of the magnetized disk expected within the MAD-accretion scenario is significantly (by several orders!) lower than the estimate of the angular velocity of the flow in traditional scenarios. Even in comparison with a quasi-spherical flow (not to mention a Keplerian disk), the rotation of the magnetized disk turns out to be extremely slow. The reason for this is the magnetic field of the flow itself, which ensures effective dissipation of the rotational energy of the infalling gas starting from the radius $r_{\text{ml}}$, at which the magnetic pressure in the flow reaches its dynamic pressure. This quality of the magnetized disk opens up prospects for using the MAD-accretion scenario in constructing a model of long-period pulsars. In this case, the equilibrium rotation of the neutron star in these objects can be explained without invoking the assumptions of a super-strong magnetic field of a neutron star or intensive dissipation of the angular momentum of the accretion flow at the magnetosphere boundary. The latter circumstance is an essential condition of all models of the rotational evolution of neutron stars based on the scenarios of the accretion from a Keplerian disk or quasi-spherical accretion. It is obvious that the assumption about the high-efficient dissipation of the angular momentum at the magnetosphere boundary is necessary in this situation, i.e., when the angular velocity of the flow at the boundary of the neutron star magnetosphere significantly exceeds the angular velocity of the neutron star itself (see Table \ref{tab:angular}). Otherwise, the equation of equilibrium rotation of the neutron star (see expression (\ref{eq:equilibrium})) has no real roots and, thus, no physical solution. In contrast, the effect of dissipation of the angular momentum of the accretion flow, arising due to its interaction with the magnetic field of the neutron star, in the MAD-accretion scenario turns out to be a second-order effect. Equation (\ref{eq:equilibrium}) in this case admits a physical solution even in the non-dissipative approximation. The possibility of taking into account dissipative effects and estimating corrections to the parameters of our model will be considered in a subsequent paper.

\section{Conclusions}
The presented estimates indicate the possibility of implementing the MAD-accretion scenario in the HMXBs we considered. The observed periods of the accreting pulsars in these systems correspond to the equilibrium period of rotation of a neutron star expected in the MAD-accretion scenario under the condition of significant magnetization and high temperature of the stellar wind of the massive component. The value of the equilibrium period of the pulsar within the framework of our scenario is determined by the degree of magnetization of the stellar wind of the massive component of the system and, other things being equal, increases with the decrease of the parameter $\beta_0$.

\section*{Acknowledgments}
The authors express their gratitude to the reviewer for useful comments.

\section*{Funding}
The work was carried out with the financial support of the Ministry of Science and Higher Education of the Russian Federation, grant No. 075-15-2022-262 (13.MNPMU.21.0003), within the framework of the research program to variable stars at the 6-m telescope of SAO RAS (BTA).

\section*{Conflict Of Interest}
The authors of this work declare that they have no conflicts of interest.

\bigskip
\begin{flushright}
{\it Translated by N. Oborina.}
\end{flushright}

\begin{thebibliography}{99}
\bibitem[Beskovnaya \& Ikhsanov(2024)]{beskrovnaya2024} Beskrovnaya, N. G., \& Ikhsanov, N. R. 2024, \textit{Astrophysical Bulletin}, 79(1), 104

\bibitem[Bisnovatyi-Kogan(2019)]{bisnovatyi2019} Bisnovatyi-Kogan, G. S. 2019, \textit{Universe}, 5(6), 146

\bibitem[Bisnovatyi-Kogan \& Fridman(1970)]{bisnovatyi1970} Bisnovatyi-Kogan, G. S., \& Fridman, A. M. 1970, \textit{Soviet Astronomy}, 13, 566

\bibitem[Bisnovatyi-Kogan \& Ruzmaikin(1974)]{bisnovatyi1974} Bisnovatyi-Kogan, G. S., \& Ruzmaikin, A. A. 1974, \textit{Astrophysics and Space Science}, 28(1), 45

\bibitem[Bisnovatyi-Kogan \& Ruzmaikin(1976)]{bisnovatyi1976} Bisnovatyi-Kogan, G. S., \& Ruzmaikin, A. A. 1976, \textit{Astrophysics and Space Science}, 42(2), 401

\bibitem[Cox et al.(2005)]{cox2005} Cox, N. L. J., Kaper, L., \& Mokiem, M. R. 2005, \textit{A\&A}, 436(2), 661

\bibitem[Delgado-Martí et al.(2001)]{delgado2001} Delgado-Martí, H., Levine, A. M., Pfahl, E., \& Rappaport, S. A. 2001, \textit{ApJ}, 546(1), 455

\bibitem[Falanga et al.(2015)]{falanga2015} Falanga, M., Bozzo, E., Lutovinov, A., et al. 2015, \textit{A\&A}, 577, A130

\bibitem[Finger et al.(2010)]{finger2010} Finger, M. H., Ikhsanov, N. R., Wilson-Hodge, C. A., \& Patel, S. K. 2010, \textit{ApJ}, 709(2), 1249

\bibitem[Hambaryan et al.(2022)]{hambaryan2022} Hambaryan, V., Stoyanov, K. A., Mugrauer, M., et al. 2022, \textit{MNRAS}, 511(3), 4123

\bibitem[Igumenshchev et al.(2003)]{igumenshchev2003} Igumenshchev, I. V., Narayan, R., \& Abramowicz, M. A. 2003, \textit{ApJ}, 592(2), 1042

\bibitem[Ikhsanov(2007)]{ikhsanov2007} Ikhsanov, N. R. 2007, \textit{MNRAS}, 375(2), 698

\bibitem[Ikhsanov \& Beskrovnaya(2012)]{ikhsanov2012} Ikhsanov, N. R., \& Beskrovnaya, N. G. 2012, \textit{Astronomy Reports}, 56(8), 589

\bibitem[Ikhsanov \& Beskrovnaya(2013)]{ikhsanov2013} Ikhsanov, N. R., \& Beskrovnaya, N. G. 2013, \textit{Astronomy Reports}, 57(4), 287

\bibitem[Ikhsanov et al.(2024)]{ikhsanov2024} Ikhsanov, N. R., Kim, V. Y., \& Beskrovnaya, N. G. 2024, \textit{Publications of the Pulkovo Observatory}, 233, 34

\bibitem[Ikhsanov \& Mereghetti(2015)]{ikhsanov2015} Ikhsanov, N. R., \& Mereghetti, S. 2015, \textit{MNRAS}, 454(4), 3760

\bibitem[Kim et al.(2023)]{kim2023} Kim, V., Izmailova, I., \& Aimuratov, Y. 2023, \textit{ApJS}, 268(1), 21

\bibitem[Klus et al.(2014)]{klus2014} Klus, H., Ho, W. C. G., Coe, M. J., et al. 2014, \textit{MNRAS}, 437(4), 3863

\bibitem[Narayan et al.(2003)]{narayan2003} Narayan, R., Igumenshchev, I. V., \& Abramowicz, M. A. 2003, \textit{PASJ}, 55, L69

\bibitem[Ruifert(1999)]{ruifert1999} Ruifert, M. 1999, \textit{A\&A}, 346, 861

\bibitem[Schulz et al.(2003)]{schulz2003} Schulz, N. S., Canizares, C., Huenemoerder, D., \& Tibbeis, K. 2003, \textit{ApJ}, 595(1), 365

\bibitem[Shakura(1975)]{shakura1975} Shakura, N. I. 1975, \textit{Soviet Astronomy Letters}, 1, 223

\bibitem[Shvartsman(1971)]{shvartsman1971} Shvartsman, V. F. 1971, \textit{Soviet Astronomy}, 15, 377
\end{thebibliography}
\end{document}